\documentclass[runningheads]{llncs}
\usepackage[T1]{fontenc}
\usepackage{graphicx}
\begin{document}
\title{A Treatment of EIP-1559: Enhancing Transaction Fee Mechanism through $N^{th}$-Price Auction}
\titlerunning{Abbreviated paper title}
% If the paper title is too long for the running head, you can set
% an abbreviated paper title here

\author{Kun Li \inst{1}%\orcidID{0000-0002-8305-0841} 
\and
Guangpeng Qi  \inst{2} 
\and
Guangyong Shang\inst{2} 
\and
Wanli Deng \inst{1}
\and
Minghui Xu\inst{1*}%\orcidID{0000-0003-3675-3461} 
\and
Xiuzhen Cheng\inst{1}%\orcidID{0000-0001-5912-4647}
}
\authorrunning{K. Li et al.}
% First names are abbreviated in the running head.
% If there are more than two authors, 'et al.' is used.
%
\institute{Shandong University, Jinan, China\\
\email{kunli@sdu.edu.cn, dengwanli7@gmail.com, \{mhxu, xzcheng\}@sdu.edu.cn,}\\
\and
Inspur Yunzhou Industrial Internet Co., Ltd., Jinan, China\\
\email{\{qigp, shangguangyong\}@inspur.com}\\
* Corresponding author: Minghui Xu
}
\maketitle              % typeset the header of the contribution
\begin{abstract}
    With the widespread adoption of blockchain technology, the transaction fee mechanism (TFM) in blockchain systems has become a prominent research topic. An ideal TFM should satisfy user incentive compatibility (UIC), miner incentive compatibility (MIC), and miner-user side contract proofness ($c$-SCP). However, state-of-the-art works either fail to meet these three properties simultaneously or only satisfy them under certain conditions. In this paper, we propose a burning $N$-price auction TFM named BNP. This mechanism divides the transaction fee into a base fee, which is burned, and a priority fee, which is allocated to miners. Theoretical proofs and experimental analyses demonstrate that, even under conditions of significant transaction congestion, this mechanism satisfies UIC, MIC, and $c$-SCP simultaneously. Furthermore, the BNP mechanism is not constrained by the type of blockchain consensus, making it widely applicable.
\end{abstract}

\section{Introduction}\label{sec:intro}
% 研究背景和意义
The improvement of hardware performance, the widespread embrace of cryptocurrencies, and ongoing refinements in consensus mechanisms have propelled the extensive utilization of blockchain technology across diverse transactional scenarios, including data storage and sharing\cite{xu2024filedes,guo2024bft,guo2023filedag}, data asset trading\cite{dai2019sdte,zheng2020blockchain}, distributed learning\cite{xu2022spdl,wang2023incentive}, among others. This diverse range of application domains has resulted in an increase in the variety and complexity of transactions within blockchain systems. To expedite transaction validation and maintain the stability and integrity of blockchain networks, transaction fee mechanisms (TFM) have been introduced, representing the fees users must pay for on-chain transactions. 

According to \cite{chung2024collusionresilience}, an ideal Transaction Fee Mechanism (TFM) should satisfy the following three properties:
\begin{enumerate}
    \item User Incentive Compatible (UIC): When other participants in the system behave honestly, users will bid truthfully.
    \item Miner Incentive Compatible (MIC): When other participants in the system behave honestly, miners will adhere to the rules established by the TFM.
    \item Miner-user Side Contract Proofness ($c$-SCP): Collusion between miners and up to $c$ users cannot increase their joint payoff through dishonest behavior.    
\end{enumerate}

In some blockchain systems, such as Bitcoin\cite{nakamoto2008bitcoin}, all transaction fees submitted by users belong to the miners. But this particular type of TFM fails to meet the aforementioned three properties, thereby presenting several drawbacks. Firstly, it lacks fairness, as miners tend to prioritize transactions with higher fees, making it challenging for lower-fee transactions to be included in blocks. Secondly, users may experience suboptimal transaction experiences due to the unpredictability of miners' behavior, leading to overpayment or transaction delays. 

To address these issues, Ethereum's EIP-1559 proposal\cite{EIP1559} has introduced a complex TFM that includes a fixed-per-block fee, which is burned. This innovative approach aims to enhance fee transparency, predictability, and fairness, ultimately improving the overall efficiency and user experience of the blockchain network. 

%ensure alignment of incentives not only for individual users but also for the miners responsible for validating blocks. Moreover, it should be designed to withstand collusion between miners and users, thereby safeguarding the integrity and fairness of the system. 
However, TFMs based on second-price auctions, such as EIP-1559, often struggle to satisfy all three properties simultaneously under the constraint of limited block size \cite{roughgarden2023transaction}. For instance, during congestion on the Ethereum network, EIP-1559 may resort to violating user incentive compatibility by reverting to first-price auctions. Several studies explore alternative approaches to address this challenge in specific scenarios or consensus mechanisms. One notable study, \cite{tang2023transaction}, examines a TFM based on the burning second-price auction for Proof of Stake (PoS) and demonstrates its ability to satisfy all three conditions under certain circumstances. Motivated by this research, we propose BNP mechanism —— a TFM for congested states based on the burning $N^{th}$-price auction. Through a combination of theoretical analysis and experimental validation, we demonstrate that this mechanism fulfills all three conditions under the Proof of Work (PoW) consensus. The contributions in our paper are summarized as follows:

%这里看后续能否拿到以太坊拥堵的数据，如果能就放上
% 本文贡献
\begin{enumerate}
    \item We proposed a TFM in blockchain system that is not restricted by the type of blockchain consensus, thus enabling its application in a wider range of scenarios.
    \item The mechanism, through the combination of a burning mechanism and an N-th price auction, ensures compliance with UIC, MIC, and c-SCP even under conditions of significant transaction congestion, thus addressing the shortcomings of EIP-1559.
    \item Theoretical analysis and experimental verification both verify the effectiveness of the proposed mechanism.
\end{enumerate}

% 本文结构
The rest of this paper proceeds as follows. In Section \ref{sec:rw}, we summarize the related work on TFM in blockchain. An overview of our proposed TFM is presented in Section \ref{sec:overview}, which is specifically elaborated in Section \ref{sec:TFM}. In Section \ref{sec:experiment}, we conduct substantial  experiments to evaluate our proposed mechanism. Finally, we conclude the whole paper in Section \ref{sec:conclusion}.

%以太坊网络上运行着种类繁多的应用程序，时刻需要处理庞杂的交易，包括 DeFi、隐私混合器和在线市场的原子交换。这些交易大多具有不同的紧急性，例如，DeFi 交易需要更快处理，而第二层支付系统的交易则相对较稳定。及时处理紧急对用户的价值和利益影响巨大，而降低非紧急交易的费用能使网络更活跃健康。

%本质上，以太坊系统可以被视为等待确认的交易排队系统[8-11]。一笔交易的成功发出首先由用户确定每笔交易内容，附上交易费后提交交易到内存池，池中的交易均在未确认状态等待矿工的挑选，由于区块大小有限，可记录到区块中的交易数受限，矿工优先选择交易费率较高的交易打包，计算寻找符合条件的nonce，最后发布到p2p网络。

%从用户的角度，交易中提供的交易费直接决定了这笔交易被矿工确认的优先级，通常只占转账的一小部分；然而在小额转账中，这一情况可能不再成立——随着网络中单位时间交易不断增多，在区块大小一定的情况下，交易被确认的时间会不断变大，大量具有最高紧急性和价值的交易同时涌入将推高交易费，导致内存池中出现大量交易拥塞，甚至达到低紧急性交易的交易费变得过于昂贵而无法使用的程度（止步、中途退出）。在这种情况下，用户需要在等待成本和交易费cost之间作出权衡，制定个性化的策略。我们可以看到在整个交易被确认的过程中，交易费用对参与者的个人决策乃至系统层面的政策制定或机制设计都有很大的影响。

%一个交易系统应该对所有使用者具备良好的激励兼容。在挖矿阶段，最先解决计算难题的矿工将获得奖励，这笔奖励在非繁忙阶段主要是新的区块奖励，因为绝大部分base fee会被燃烧掉，以确保机制是UIC的。而在网络出现拥塞时，大部分交易携带priority fee使得不再满足UIC。而【1】之前的所有工作都未能同时实现所有三个属性UIC、MIC、oca。The closest we have come to achieving all three properties is Ethereum’s recent EIP-1559 [4] proposal.  但Roughgarden [31, 32] 认为，当出现拥塞时，EIP-1559 的行为类似于最高价拍卖，因此无法满足 UIC，即策略性竞价可以提高单个用户的效用。

%尽管有论文[2]指出不存在同时满足UIC和1-SCP的价值TFM。但文章都是针对特定不变的机制，而对于能适配交易特性的动态机制没有论述。而我们的动态机制做到了这一点。

\section{Related Works}\label{sec:rw}
As blockchain technology continues to gain traction across diverse industries and applications, understanding and optimizing transaction fee mechanisms have become paramount for enhancing the efficiency, fairness, and stability of blockchain ecosystems. Consequently, the study of transaction fees within blockchain systems has emerged as a crucial area of research. 

Liu et al.\cite{61e7815b5244ab9dcbf99cf3}conducted a systematic evaluation of the real-world impacts of TFM. They examined the causal effects of EIP-1559 on blockchain transaction fee dynamics, transaction wait times, and consensus security, proposing new directions for improving TFM based on their findings. Liu et al. \cite{6049e94691e01118b758f044} addressed the issue of blockchain storage sustainability. They proposed that transaction fees could help offset the increasing storage costs for miners and designed a social welfare-maximizing mechanism. This mechanism models the interaction between protocol designers, users, and miners as a three-stage Stackelberg game, incentivizing each user to pay sufficient transaction fees to cover storage costs.

In contrast to the focus of the aforementioned studies, Landis et al. \cite{6456389bd68f896efacf6966} explored Stackelberg attacks on the transaction fee auction process. These attacks are applicable to first-price auctions, second-price auctions, and the transaction fee mechanism used in Ethereum's EIP-1559. Their research highlights the critical importance of designing transaction fees that are incentive-compatible to prevent such vulnerabilities.

Unlike the general area of mechanism design, which primarily focuses on bidder strategies and behavior, blockchain transaction fee auctions involve miners' behavior, who, as auctioneers, have incentives to engage in malicious behavior\cite{chen2022bayesian,wu2023maximizing}. Building on a detailed analysis of the characteristics of blockchain transaction scenarios, \cite{chung2022foundations,lavi2022redesigning,roughgarden2023transaction} propose that an ideal TFM should satisfy three key properties: user incentive compatibility, miner incentive compatibility, and miner-user side contract proofness. 

Tang et al.\cite{tang2023transaction} extend the TFM in \cite{chung2022foundations} by incorporating a long-run utility model for the miner. Their $BSP(\theta)$ mechanism is shown to satisfy all three desired properties of a TFM when the parameter $\theta$ falls within a specific range and an appropriate “tick” size is imposed on user bids.

Chung et al.\cite{chung2024collusionresilience} prove that when there is contention between transactions, no (possibly randomized) direct-revelation TFM can satisfy UIC, MIC, and OCA-proofness (off-chain agreement, which refers to colluding strategies between the miner and a set of users that allow off-chain transfers). They also explore possible ways to circumvent these impossibilities.

Building on these studies, we propose the BNP mechanism, which combines the burning mechanism and $N$-price auction. This mechanism takes into account the long-run payoff of both miners and users and is proved to satisfy the three properties of the ideal TFM without being constrained by the type of consensus.

\section{Preliminary and Overview}\label{sec:overview}
According to the analysis in Section \ref{sec:intro}, Ethereum proposed EIP-1559\cite{EIP1559} which introduces a significant improvement by incorporating a burning mechanism. This improvement aims to simultaneously satisfy the three properties of TFM: user- and miner-incentive compatibility (UIC and MIC) as well as c-SCP, under the assumption that the block size is unlimited and there is no congestion in the blockchain system. Specifically, EIP-1559 specifies a base fee for each block, and all transactions within that block will burn an amount of ETH equal to base fee multiplied by gasUsed. In addition to the base fee, users have the option to increase the priority fee to prioritize their transactions for inclusion in the next block. This smoothes out the fluctuations in transaction fees and reduces the need for user intervention in transaction sequencing. However, in reality, block sizes are limited, and blockchain transaction systems are inevitably prone to congestion. This results in the degradation of the TFM based on base fee \& priority fee into a first-price auction centered around the priority fee, thereby no longer being able to simultaneously satisfy the properties of TFM. 

Building upon the foundations laid in \cite{chung2022foundations}, \cite{tang2023transaction} introduced the burning second-price auction mechanism $BSP(\theta)$ for PoS consensus. This mechanism, under the condition that the block size is limited, has been demonstrated to satisfy all three desired properties of the TFM under specified conditions. 

\begin{figure}[htbp]
	\centering
	\includegraphics[width=0.75\textwidth]{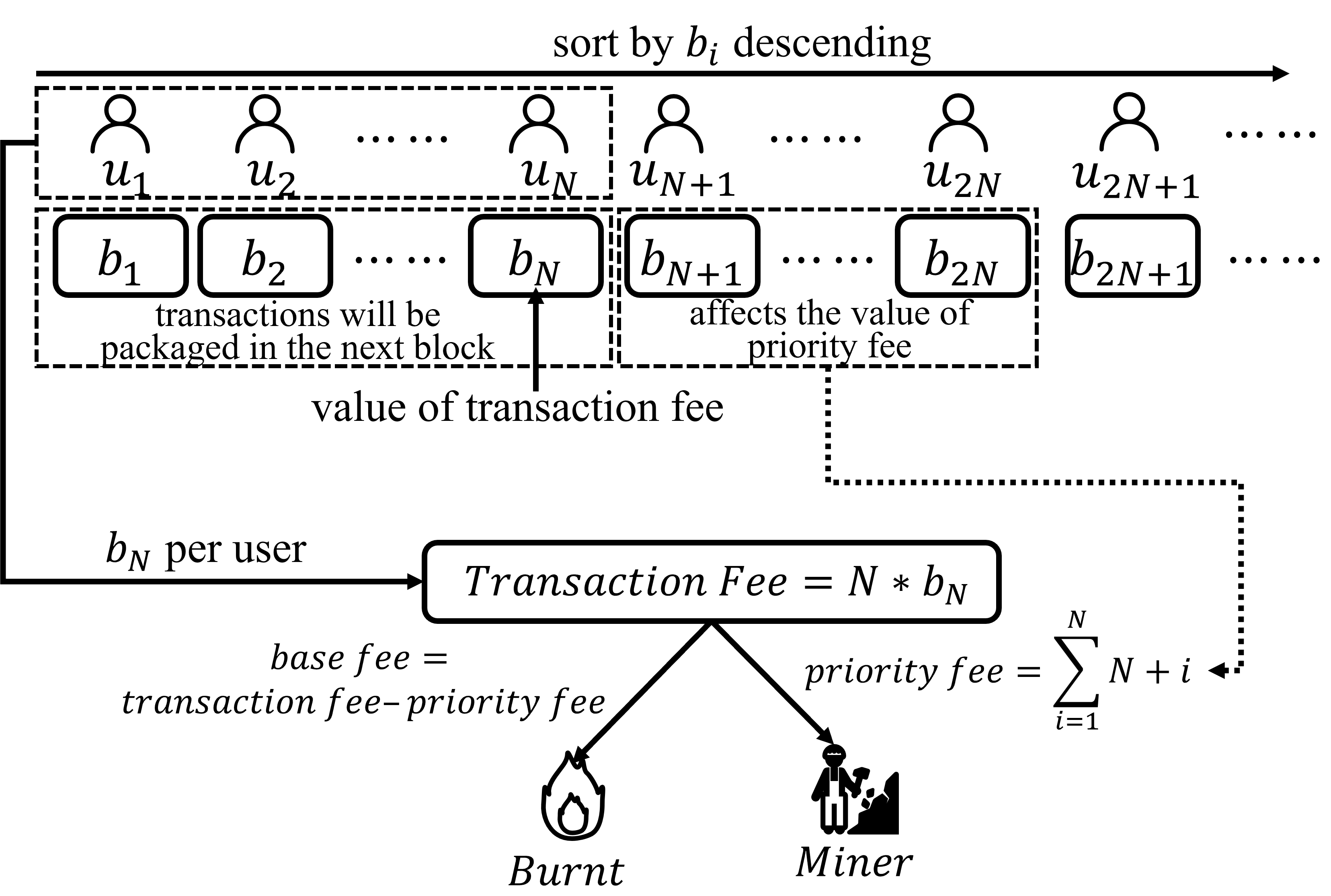}
	\caption{Overview of BNP mechanism.}
	\label{fig:overview}
\end{figure}

Inspired by this research, we propose the $BNP$ mechanism as illustrated in Fig. \ref{fig:overview}. In this mechanism, all users submit their bids for transaction fees in the form of a sealed-bid auction, and the top $N$ bidders will pay at the $N^{th}$ price. Similar to EIP-1559, this mechanism also divides the transaction fee into base fee and priority fee, where the base fee is burned, and the priority fee goes to the miners.

Specifically, transactions awaiting packaging in the transaction pool are sorted by bid for transaction fee from high to low, denoted as $\{b_1, b_2, ..., b_N, b_{N+1},...\}$, where $N$ represents the number of transactions to be included in the next block. Following the principles of the $N$-price auction mechanism, the first $N$ transactions will pay at the $N^{th}$ transaction's bid price. That is, transactions $\{t_1, t_2, ..., t_N\}$ will be included in the next block with the cost of $b_N$. $b_i-b_N, i \in \{1,2,...,N\}$ will be refunded to user $u_i$. Of the total transaction fees $N \times b_N$ submitted to the blockchain system, $\sum_{i=1}^N b_{i+N}$ is considered the priority fee, which goes to the miners, while $N \times b_N - \sum_{1=1}^N b_{i+N}$ is considered the base fee, which will be burned. If the number of transactions awaiting packaging is less than $2N$, the missing values will be filled with zeros.

Unlike \cite{tang2023transaction}, the mechanism is not specific to any particular type of consensus. In the following discussion, we will demonstrate how this mechanism can still satisfy the three properties of TFM: UIC, MIC, and c-SCP even under conditions of limited block capacity and high transaction volume congestion.

\section{The Properties of BNP Mechanism}\label{sec:TFM}
From the above content, it is evident that when all users' bids for transaction fees are arranged in descending order as ${b_1, b_2, ..., b_N, b_{N+1},...}$, the miner's payoff function is:
\begin{equation}\label{eq:pm}
    \mathcal{P}_m = \sum_{i=1}^N b_{i+N}
\end{equation}

In a given round of auctions, if a user $u_i$'s transaction $t_i$ is successfully included in the next block with the cost of $b_N$, he will receive the revenue generated from having the transaction confirmed on-chain. It's evident that $u_i$'s bid $b_i$ for the transaction fee is directly related to this revenue and cannot exceed it. Each user $u_i$ has a psychological price $b_i^{true}$ for the transaction fee, indicating the fee they are willing to pay. If the actual transaction fee paid exceeds $b_i^{true}$, $u_i$ will perceive it as a loss. For simplicity, we assumes that the revenue $u_i$ can generate from having his transaction included in a block is equivalent to $b_i^{true}$. In summary, if $u_i$'s transaction is successfully included in the next block, he will receive $b_i^{true}$ with a cost of $b_N$, resulting in the user's payoff function is:
\begin{equation}\label{eq:pu}
    \mathcal{P}_u=b_i^{true}-b_N
\end{equation}

An honest user's bid $b_i$ equals $b_i^{true}$. Clearly, in scenarios with a high volume of transactions and congested trading systems, users have the motivation to engage in the following dishonest behaviors which undermines UIC:
\begin{enumerate}
    \item Fake bid: Refers to the user fabricating a transaction and submitting a bid to participate in the auction for transaction fees, despite having no actual transaction need.
    \item Overbid: Refers to the user submitting a bid that exceeds his psychological price, i.e., $b_i > b_i^{true}$.
    \item Underbid: Refers to the user submitting a bid that is lower than his psychological price, i.e., $b_i < b_i^{true}$.
\end{enumerate}

On the other hand, in pursuit of higher profits, miners also have an incentive to fabricate a bid higher than $b_{N}$, intending to artificially inflate the clearing price without being detected, which renders MIC unfulfilled. It's important to note that due to the transparency of the blockchain transaction system, the transaction corresponding to this fake bid needs to be genuinely sent out by the miner and added to the pending transaction pool. If this transaction is not included in the current miner's block, it's highly likely to be included by other miners in the future. This implies that the miner would need to pay transaction fees to other miners.

Furthermore, since a miner's revenue is directly determined by bids from the $N+1^{th}$ to the $N+N^{th}$ positions, miners may also collude with users through off-chain negotiations to conspire. This collusion could involve incentivizing certain users to increase their bids, thus allowing the miner to achieve higher profits, which contradicts the requirement of $c$-SCP. 

An ideal TFM should eliminate the incentives for all participants to engage in the aforementioned malicious behaviors. In the following sections, we will systematically demonstrate how the mechanism proposed in this paper, BNP, can probabilistically satisfy the three properties of the ideal TFM: UIC, MIC, and $c$-SCP.

%一个理想的TFM应该使得所有系统参与者都没有动机采取上述恶意行为。我们将在下文中逐一证明本文所设计的机制BNP能够大概率满足ideal TFM的三个属性，即ensure alignment of incentives not only for individual users but also for the miners responsible for validating blocks. Moreover, it is also designed to withstand collusion between miners and users, thereby safeguarding the integrity and fairness of the system.

%在本轮竞拍中，用户$u_i$所发出的交易$t_i$如果能成功上链，则其cost是$b_N$，并会获得上链所能取得的收益。显然$u_i$的报价bid $b_i$与该收益正相关，并且不会超过该收益。为了简化分析，我们将该收益定义为用户真实价格$b_i^{true}$，即：用户的交易如果能在本轮上链则会获得收益$b_i^{true}$，一个诚实用户的bid $b_i=b_i^{true}$。【加引用说明交易上链收益是随时间损耗的】

% 显然，当交易量很大、交易系统拥堵的情况下，在上述交易费机制下，用户有可能会overbid、underbid或fakebid，

%另一方面，为了获取更高收益，矿工也有动机伪造一个高于$b_{N}$的bid，在不被发现的情况下抬高成交价。【这里要说明这个bid是真的要发在区块链系统中，如果这次不被矿工自己上链，则后续有可能被其他矿工上链，意味着要付出手续费】

%此外，由于矿工的收益直接由第$N+1$到第$N+N$个bid决定，因此矿工也有可能与用户合谋，通过链下交易买通某些用户，让其提高bid从而获取更高收益。

%一个理想系统的激励机制应该诱导利益相关者选择合理的行为，而不是只约束一部分参与者的行为。我们将在下文中逐一证明本文所设计的$k$价拍卖机制满足激励相容，即用户和矿工不会伪造报价和共谋。

\subsection{User Incentive Compatible}
In this section, we examine whether BNP adheres to UIC, ensuring that if $u_i$ bids truthfully ($b_i=b_i^{true}$), his expected utility is maximized, assuming all other system participants are honest.

Based on the payment rules of the blockchain system, any fake bid offers no value to the user and necessitates paying transaction fees. It's evident that without collusion, users cannot gain any profit from fake bids; on the contrary, they would incur losses. Engaging in fake bidding is inherently irrational, as rational users lack motivation to partake in such behavior. Hence, in this section, our primary focus lies in analyzing two forms of dishonest conduct by users: underbidding and overbidding.

%为了避免用户overbid、underbid或fakebid，本节关注BNP是否满足UIC，即是否能实现：在其余系统参与者诚实的前提下，如果$u_i$ bid truthfully，即$b_i=b_i^{true}$，则用户期望效用总是最大的。

%根据区块链系统支付规则，任何fake bid为用户带来的价值为0，且需要支付交易费，显然，在不考虑共谋的前提下，任何用户都无法从fake bid获取收益，反而会面临损失，fake bid是非理性行为，一个理性的用户没有动机fake bid。因此，本节中我们重点分析用户underbid和overbid这两个不诚实行为。

% TODO 加图
Let's delve into the scenario of overbidding, i.e., $b_i>b_i^{true}$. For $t_i$, $u_i$'s bid can take on the following possibilities:
\begin{enumerate}
    \item $b_N>b_i>b_i^{true}$: This indicates that the transaction will not be immediately packaged, resulting in no immediate consequences. However, there's a chance it might be packaged in the future at a price higher than $b_i^{true}$, which would result in a loss. Therefore, users will likely refrain from overbidding.
    \item $b_i\geq b_N\geq b_i^{true}$: Initially, when the bid is at $b_i^{true}$, the transaction might not qualify for packaging. However, by increasing the bid to $b_i$, the transaction could become eligible for packaging at the expense of $b_N$. Nevertheless, the profits accrued from being packaged are outweighed by the costs users incur. Consequently, users are unlikely to engage in overbidding within this scenario.
    \item $b_i>b_i^{true}\geq b_N$: In this case, the transaction can be packaged regardless of whether the user overbids. This scenario can be further divided into two subcases:
    \begin{enumerate}
        \item $b_i>b_i^{true}=b_N$: This suggests that the bid precisely occupies the $N^{th}$ position in the descending order of bids. Following an overbid, all users will be required to pay transaction fees of $\min{b_i, b_{N-1}}$. It's evident that the cost still surpasses the profit, hence users will continue to abstain from overbidding.
        \item $b_i>b_i^{true}>b_N$: This indicates that the bid ranks among the top $N-1$ bids after being sorted in descending order. Regardless of whether the user decides to overbid, the final transaction fee required remains $b_N$, and the user's payoff remains unchanged. Consequently, users lack any incentive to engage in overbidding within this scenario.
    \end{enumerate}
\end{enumerate}

In summary, users refrain from overbidding, considering the possibility of transactions being packaged and the associated costs.

Next, let's delve into the scenario of underbidding, where $b_i < b_i^{true}$. This scenario can be categorized as follows:
\begin{enumerate}
    \item $b_N>b_i^{true}>b_i$: Regardless of whether the user underbids, the transaction will not be packaged. Additionally, since the new bid is lower, miners tend to prioritize higher bids, potentially delaying the packaging of this transaction. Consequently, users lack motivation to underbid in this case.
    \item $b_i^{true}\geq b_N>b_i$: In this case, if the user refrains from underbidding, the transaction is initially eligible for packaging. However, once underbidding occurs, the situation requires further examination, leading to the following two scenarios:
    \begin{enumerate}
        \item $b_i^{true}\geq b_N>b_{N+1}\geq b_i$: After underbidding, the transaction cannot be packaged, resulting in no corresponding profit. Similarly, due to the lower bid, the transaction's packaging might be delayed. Therefore, users lack motivation to underbid.
        \item $b_i^{true}\geq b_N>b_i>b_{N+1}$: After underbidding, the transaction can still be packaged, and the transaction fee for all users decreases from $b_N$ to $b_i$. In this case, since expenses decrease, users have motivation to underbid.\label{underbid}
    \end{enumerate}    
    \item $b_i^{true}>b_i\geq b_N$: Regardless of whether the user underbids, the transaction will be packaged with the cost of $b_N$. Since the user's payoff remains unchanged, there is no motivation for the user to underbid.
\end{enumerate}

Since transaction fees operate on a sealed-bid auction basis, users cannot predict the specific outcome of underbidding when making their bids. Moreover, during congestion, the difference between $b_N$ and $b_{N+1}$ tends to be minimal, indicating a low probability of Scenario \ref{underbid}. Additionally, the potential gains are limited, not exceeding $b_N-b_{N+1}$. Without knowledge of other bids, underbidding poses a significant risk for relatively small returns. Considering all scenarios, rational users are unlikely to underbid.
%\added{add figure here} %【这里需要加个图或数据分析，说明拥堵时确实是这样的】

Based on the above analysis, it can be concluded that the BNP mechanism has a high probability of satisfying UIC.

\subsection{Miner Incentive Compatible}\label{sec:MIC}
This section analyzes whether the miner can achieve higher profits through fake bids. 

Unlike users, the miner have access to all bids, allowing him to precisely evaluate the potential profits from various fake bids and choose the most lucrative option. We denote the fake bid chosen by the miner as $b_{fake}$. As previously discussed, in the BNP mechanism, user expenses are determined by the $N^{th}$ bid, while miner profits are determined by bids ranging from the $(N+1)^{th}$ to the $2N^{th}$ bid. Clearly, the miner would not choose to submit a bid lower than $b_{2N}$, as it would have no impact on his profits and would instead entail paying transaction fees for that transaction in the future. Moreover, we can further analyze the potential fake bids that the miner may consider in the following scenarios:
\begin{enumerate}
    \item $b_N>b_{fake}\geq b_{2N}$: In this scenario, the fake bid submitted by the miner would displace $b_{2N}$ from the previous top $2N$ bids. The miner's payoff becomes $\mathcal{P}_m=\sum{i=1}^{N-1} b_{i+N}+b_{fake}$. Comparing this with Equation \ref{eq:pm}, it's evident that the fake bid increases the miner's profit by $b_{fake}-b_{2N}$. Clearly, to maximize profit, the miner would seek to increase $b_{fake}$ as much as possible without exceeding $b_N$, namely $b_{fake}\to b_N$. However, as previously mentioned, this transaction will be packaged by other miners in the future, leading the miner to pay a transaction fee denoted as $b_N^\prime$. In other words, the profit brought by the miner's fake bid is $\mathcal{P}^{fake}_m = b_{fake}-b_{2N}-b_N^\prime$. If no new transactions enter the system before the next block is formed, the new $(N+1)^{th}$ bid will be $b_{2N}$. Conversely, if any new transaction enters the system, and if its bid is higher, it will further push back the ranking of $b_{2N}$. Consequently, $b_N^\prime>b_{2N}$ can be deduced. Employing the inequality manipulation further yields $\mathcal{P}^{fake}_m<b_N-2b_{2N}$. During congestion, the difference between $b_N$ and $b_{2N}$ is often small, making $b_N<2b_{2N}$ likely to hold. This implies that $\mathcal{P}^{fake}_m$ is likely to be less than 0, meaning that in the long run, miners are unlikely to gain positive profits from engaging in fake bidding behavior.
    \item $b_{N-1}\geq b_{fake}\geq b_{N}$: In this scenario, the fake bid submitted by the miner displaces $b_{N}$ from the previous top $N$ bids, resulting in users needing to pay a transaction fee of $b_{fake}$ instead of $b_N$. Simultaneously, the miner must also pay $b_{fake}$ as a transaction fee for this fake bid. As $b_{N}$ becomes the $N+1^{th}$ bid in the sequence of all bids, the miner's payoff function becomes $(\sum_{0}^{N-1} b_{N+i})-b_{fake}$. Comparing this with Equation \ref{eq:pm}, it's evident that the profit brought by the fake bid for the miner is $\mathcal{P}^{fake}_m = b_N-b_{2N}-b_{fake}$. Since $b_N\leq b_{fake}$, $\mathcal{P}^{fake}_m\leq 0$, indicating that the miner cannot gain positive profits from engaging in fake bidding behavior.
    \item $b_{fake}>b_{N-1}>b_{N}$: Similarly to the previous scenario, the transaction fee for users shifts to $b_{N-1}$, while the miner's payoff function transforms into $(\sum_{0}^{N-1} b_{N+i})-b_{N-1}$. Consequently, it can be further inferred that the profit yielded by the fake bid for the miner is $\mathcal{P}^{fake}_m = b_N-b_{2N}-b_{N-1} \leq 0$. Clearly, the miner also cannot generate positive profits in this situation.
\end{enumerate}

Building on the earlier examination, it's apparent that the BNP mechanism stands a good chance of meeting the conditions for MIC.

\subsection{Miner-user Side Contract Proofness}
Finally, this section analyzes whether the BNP mechanism can satisfy the condition of $c$-SCP.

The primary distinction between collusion and miners engaging in fake bidding individually lies in the nature of the action. In the latter case, miners fabricate a transaction when there is no actual demand, resulting in their expenditure being the full transaction fee associated with the fake bid. However, in the case of collusion, miners seek out users with transaction demands and negotiate to increase the bid of such users. This implies that, following collusion, the expenditure of the coalition comprising miners and users is the difference between the transaction fees after collusion and those before collusion. The reduction in expenditure incentivizes miners to collude with users more actively.

Following the principle of starting from simplicity to complexity, we first analyze the scenario where a miner colludes with one user, known as 1-Collusion. In this scenario, a user increases their bid $b_i$ to $b_i^c$, potentially benefiting the miner, and the two parties negotiate the distribution of profits. We consider the user and the miner as a coalition and analyze their collective payoff. If the user's original bid $b_i\geq b_N$, collusion does not alter the miner's profits; instead, it might increase the user's expenses. Therefore, miners would not opt for collusion with such users. When $b_i< b_N$, although $t_i$ may not be packaged immediately, it will inevitably be packaged in the future with a transaction fee $b_N^\prime < b_i$. We can further infer that the payoff function of the coalition is $\mathcal{P}_c^{honest}=\sum_1^N b_{i+N}+b_i-b_N^\prime$ in the absence of collusion . To delve deeper into the potential scenarios involving $b_i$ and $b_i^c$, let's analyze each of the following cases separately:
\begin{enumerate}
    \item $b_i^c\geq b_N>b_i\geq b_{2N}$: Before collusion, $t_i$ remains unpackaged, but its bid $b_i$ impacts the miner's payoff $\mathcal{P}_m$. Post-collusion, $t_i$ will be packaged, yet its impact on $\mathcal{P}_m$ diminishes. Instead, the transaction $t_N$ corresponding to $b_N$ becomes unpackaged, and $b_N$ affects $\mathcal{P}_m$. This signifies that $u_i$ incurs a cost of $min{b_{N-1},b_i^c}$ to ensure the packaging of $t_i$, yielding a profit of $b_i$. Meanwhile, the miner's payoff transforms into $\sum_0^N b_{N+i}-b_i$. The payoff function of the collusion coalition is then expressed as $\mathcal{P}_c^{SCP-1}=\sum_0^N b_{N+i}-b_i+b_i-min\{b_{N-1},b_i^c\}$.
    \item $b_i^c\geq b_N>b_{2N}>b_i$: This also implies that $u_i$ incurs a cost of $min{b_{N-1},b_i^c}$ to ensure the packaging of $t_i$ and obtain a profit of $b_i$. The difference from the previous scenario is that since $b_i$ no longer affects $\mathcal{P}_m$, the miner's payoff changes to $\sum_0^{N-1} b_{N+i}$. The payoff function of the collusion coalition in this case is $\mathcal{P}_c^{SCP-2}=\sum_0^{N-1} b_{N+i}+b_i-min\{b_{N-1},b_i^c\}$.
    \item $b_N>b_i^c>b_i\geq b_{2N}$: Regardless of collusion, $t_i$ remains unpackaged, and both $b_i$ and $b_i^c$ influence $\mathcal{P}_m$. Consequently, after collusion, the miner's payoff becomes $\sum_1^N b_{N+i}-b_i+b_i^c$. On the other hand, since $t_i$ will be packaged by other miners in the future, requiring $u_i$ to pay $b_N^\prime\leq b_i^c$. Thus, in the long run, collusion implies that the user incurs a cost of $b_N^{\prime\prime}$ in exchange for a profit of $b_i$. The payoff function of the collusion coalition in this scenario is $\mathcal{P}_c^{SCP-3}=\sum_1^N b_{N+i}-b_i+b_i^c+b_i-b_N^{\prime\prime}$.
    \item $b_N>b_i^c>b_{2N}>b_i$: Similarly, $b_i$ does not affect $\mathcal{P}_m$, while $t_i$ will be packaged in the future. Consequently, after collusion, the miner's payoff becomes $\sum_1^{N-1} b_{N+i}+b_i^c$, while the user incurs a cost of $b_N^{\prime\prime}$ in exchange for a profit of $b_i$. The payoff function of the collusion coalition in this scenario is $\mathcal{P}_c^{SCP-4}=\sum_1^{N-1} b_{N+i}+b_i^c+b_i-b_N^{\prime\prime}$.
\end{enumerate}

Upon comparison of the four scenarios outlined above, it becomes evident that $\mathcal{P}_c^{SCP-1}>\mathcal{P}_c^{SCP-2},\mathcal{P}_c^{SCP-3}>\mathcal{P}_c^{SCP-4}$ and $\mathcal{P}_c^{SCP-3}>\mathcal{P}_c^{SCP-1}$. Clearly, to maximize the coalition's revenue, the miner would opt to collude with users whose bid $b_i\geq b_{2N}$, and subsequently, elevate the user's bid to $b_i^c<b_N$ (as in case 3). Analogous to the approach in Section \ref{sec:MIC}, the miner would aspire for $b_i^c\to b_N$, thereby maximizing their profit to the fullest extent. Employing inequality manipulation, we can also deduce that $\mathcal{P}_c^{SCP-3}-\mathcal{P}_c^{honest}$ is likely to trend below 0. This suggests that over the long term, miners collaborating with a single user are improbable to attain higher payoffs through collusion.

When we broaden our analysis to scenarios involving collusion between miners and $c$ users, it becomes clear that the most effective strategy for miners is to target $c$ users whose bids fall within the range of $b_{2N}$ and $b_N$, convincing them to elevate their bids to $b_i^c \to b_N$. Likewise, given that users' transactions will likely incur future transaction fees exceeding their psychological thresholds, it follows that collusion is likely to yield a negative payoff for the coalition comprising miners and $c$ users.

Drawing from the preceding analysis, it's reasonable to infer that the BNP mechanism has a strong likelihood of fulfilling the requirements for $c$-SCP.

\section{Experiment Analysis}\label{sec:experiment}
%说明实验环境、数据来源、处理依据
%说明4张图分别的含义，包括图是为了表达什么、横纵坐标的含义、从图中能看出来什么
In this section, we further study the expenditures and revenues under different behaviours of users and miners under the BNP mechanism through experiments to verify the above theoretical analysis. The experimental environment is as follows:
\begin{itemize}
    \item Hardware environment: the memory used is Intel(R) Xeon(R) Silver 4214R CPU @ 2.40GHz, the RAM is 50GB DDR4, and the storage is 1.5TB SSD.
    \item Software environment: Ubuntu22.04LTS, Python3.8.18, Numpy1.24.4, pandas2.0.3, matplotlib3.7.5, nodejs12.22.9, mysql 8.0.36-0ubuntu0.22.04.1.
\end{itemize}

We used the official interactive interface provided by Geth, the JSON-RPC API, to capture all transactions from 7,200 blocks on the Ethereum blockchain, ranging from block 15357273 to 15364473. Additionally, we analyzed transaction pool data from block 19946372 to 19952631, spanning the period from May 25, 2024, 10:43:35 AM (UTC) to May 26, 2024, 07:42:47 AM (UTC). Since this paper focuses on the BNP mechanism's performance during periods of congestion, we further filtered out the 532 blocks from the aforementioned data that experienced transaction congestion for detailed analysis.

\begin{figure}[htbp]
    \begin{minipage}[t]{0.5\linewidth}
        \centering
	    \includegraphics[width=1\textwidth]{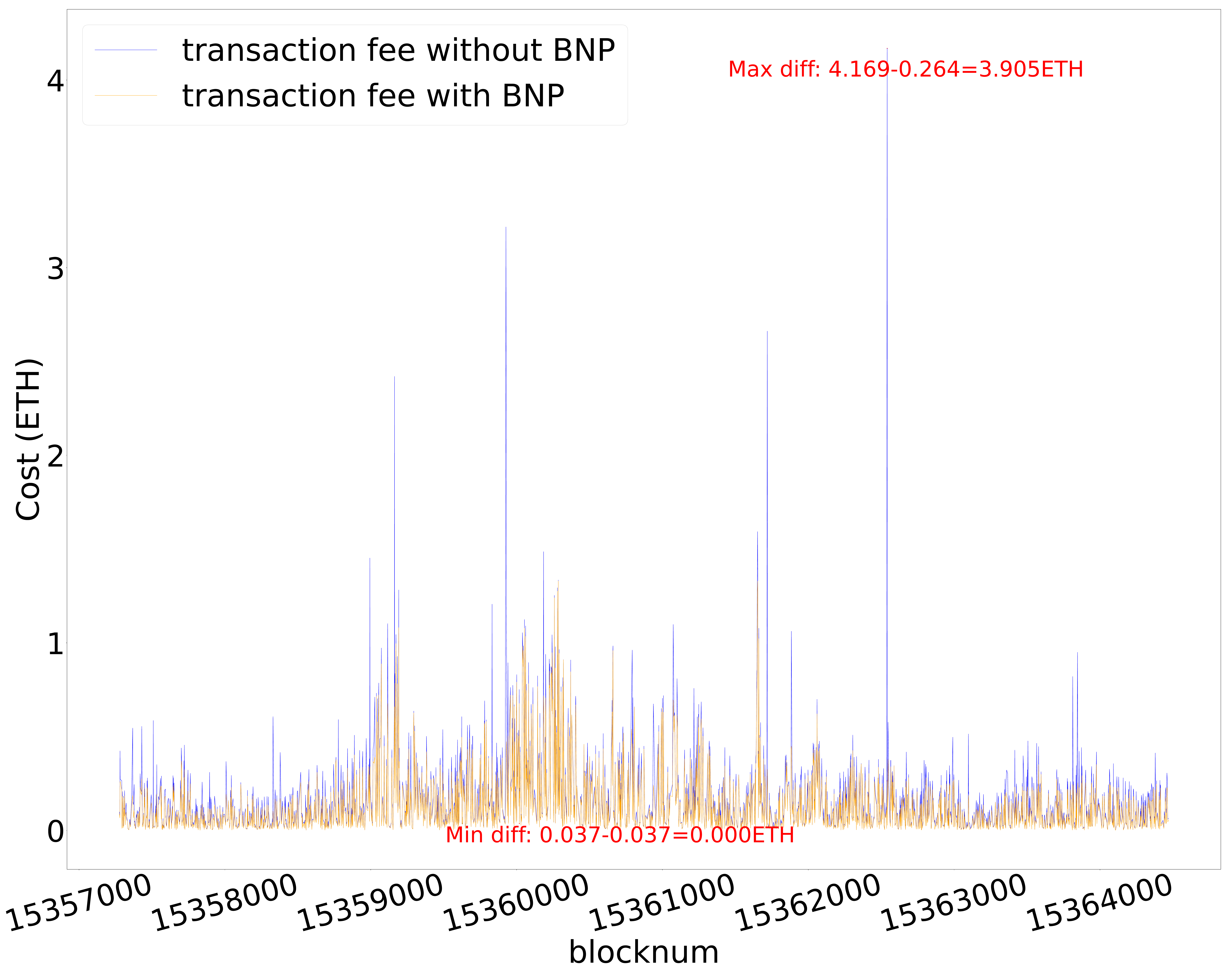}
	    \caption{User's transaction fee without BNP v.s. with BNP.}
	    \label{fig:user}
    \end{minipage}%
    \begin{minipage}[t]{0.5\linewidth}
        \centering
	    \includegraphics[width=1\textwidth]{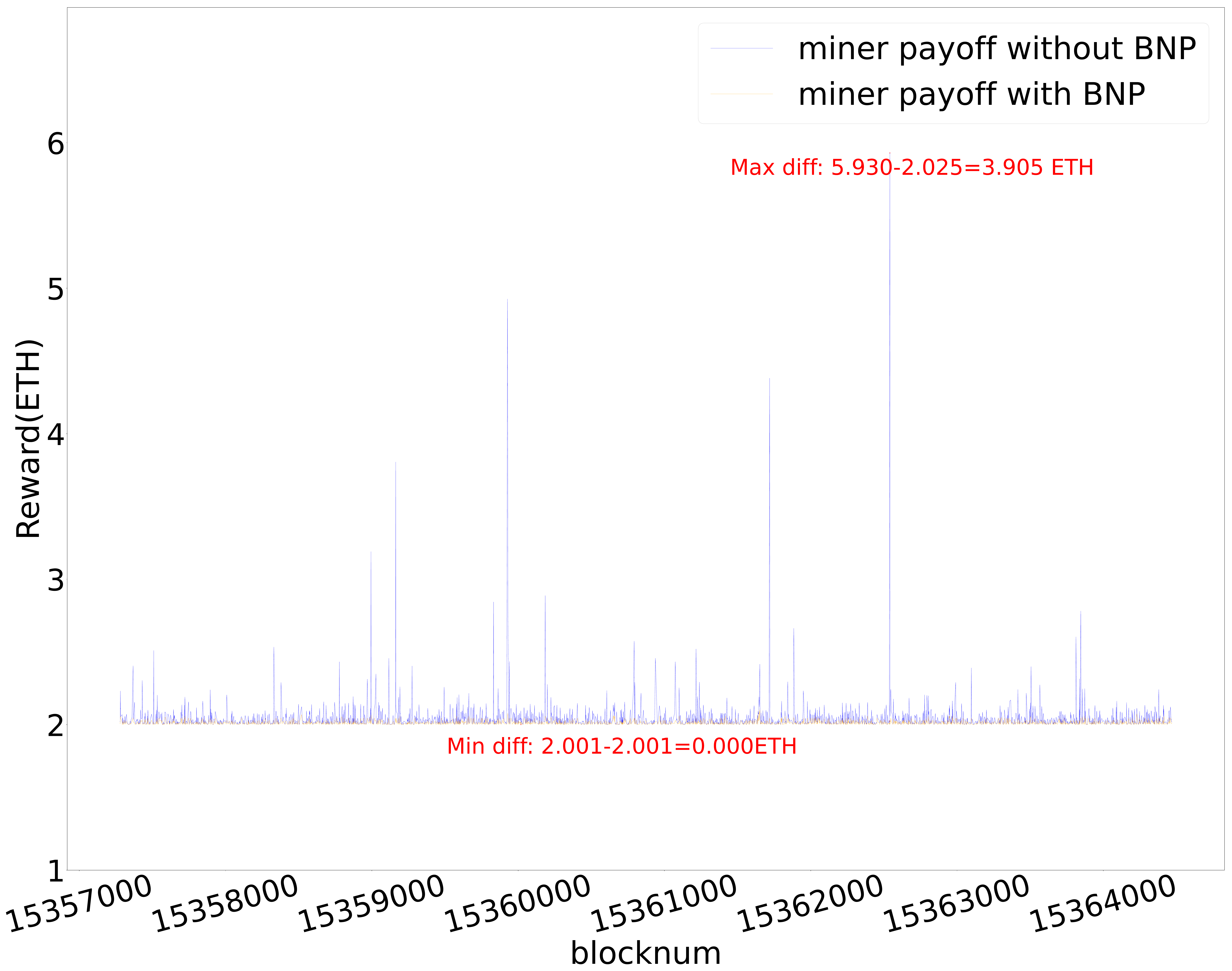}
	    \caption{Miner's payoff without BNP v.s. with BNP.}
	    \label{fig:miner}
    \end{minipage}
\end{figure}

We first analyzed the impact of the BNP mechanism on both users and miners. We applied the BNP mechanism to all transactions in 532 blocks and compared the difference in user earnings before and after implementing the mechanism. As shown in Fig. \ref{fig:user}, the use of the BNP mechanism resulted in an average reduction of 17.08\% in transaction fees. In the block with the greatest reduction, the BNP mechanism lowered the transaction fee by 3.9 ETH (a decrease of 93.75\%). In the figure, the horizontal axis represents the block number, while the vertical axis indicates the transaction fees that users need to pay in each block.

We also compared the difference in miner earnings before and after using the mechanism. As shown in Fig. \ref{fig:miner}, the BNP mechanism led to a slight reduction in miner earnings (an average decrease of 1.39\%), with the maximum reduction being 65.76\% and the minimum reduction being zero. In the figure, the horizontal axis represents the block number, while the vertical axis indicates the revenue that miners received.

From the above experiments, it can be seen that the use of the BNP mechanism significantly reduces user expenses while causing only a minimal decrease in miner earnings.

\begin{figure}[htbp]
    \begin{minipage}[t]{0.3\linewidth}
    	\centering
    	\includegraphics[width=1\textwidth]{figure/UIC.pdf}
    	\caption{The graph of payoff for users from dishonest behavior.}
    	\label{fig:UIC}
    \end{minipage}%
    \begin{minipage}[t]{0.3\linewidth}
    	\centering
    	\includegraphics[width=1\textwidth]{figure/MIC.pdf}
    	\caption{The graph of payoff for miners from dishonest behavior.}
    	\label{fig:MIC}
    \end{minipage}
    \begin{minipage}[t]{0.3\linewidth}
    	\centering
    	\includegraphics[width=1\textwidth]{figure/SCP.pdf}
    	\caption{The graph of payoff for 1-collusion coalition from dishonest behavior.}
    	\label{fig:SCP}
    \end{minipage}%
\end{figure}

Next, we verified whether the BNP mechanism satisfies UIC by randomly selecting transactions and altering their bids. As shown in Fig. \ref{fig:UIC}, in 532 blocks, only 138 blocks had users who could slightly reduce their transaction fees through dishonest behavior. In the remaining blocks, users faced higher transaction fees due to dishonest behavior. This indicates that, in the long run, the BNP mechanism satisfies UIC. In the figure, the horizontal axis represents the block number, and the vertical axis represents the change in transaction fees resulting from dishonest behavior.

Fig. \ref{fig:MIC} demonstrates that the BNP mechanism satisfies MIC. In the figure, the horizontal axis represents the block number, and the vertical axis represents the gains for miners from dishonest behavior. In 532 blocks, only 165 blocks had miners who could achieve higher payoffs through dishonest behavior. Overall, dishonest behavior led to an average reduction of 0.01 ETH in miner earnings per block. This indicates that, in the long run, the BNP mechanism satisfies MIC.

As shown in Fig. \ref{fig:SCP}, in 532 blocks, there are 208 blocks where the miner-user coalition can slightly increase its gain through dishonest behavior. In the remaining 324 blocks, the coalition has lower gains due to dishonest behavior (the final gasPrice is reduced by 0.74gWei per block on average). This shows that in the long run, the BNP mechanism satisfies 1-SCP. In the figure, the horizontal axis represents the block number and the vertical axis represents the change in transaction fees caused by dishonest behavior.

\section{Conclusion}\label{sec:conclusion}
In this paper, we propose a burning $N$-price auction TFM named BNP, which divides the transaction fee submitted by users into a base fee that is burned and a priority fee that goes to the miners. This mechanism is proven to satisfy UIC, MIC, and $c$-SCP even under conditions of transaction congestion, effectively addressing the shortcomings of EIP-1559. Experimental results demonstrate that the BNP mechanism reduces user expenses by an average of 17.8\%, while only slightly decreasing miner earnings by an average of 1.39\%. Furthermore, the BNP mechanism is not constrained by the type of blockchain consensus, making it applicable to a wider range of use cases.

%
% ---- Bibliography ----
%
% BibTeX users should specify bibliography style 'splncs04'.
% References will then be sorted and formatted in the correct style.
%
% \bibliographystyle{splncs04}
% \bibliography{mybibliography}
%
\bibliographystyle{splncs04}
\bibliography{reference}

\end{document}